\documentclass[twoside,leqno,twocolumn]{article}
\usepackage{float}
\usepackage{amsmath,bm}
\usepackage{amsfonts}
\usepackage{latexsym}
\usepackage{array}
\usepackage{graphicx}
\usepackage{url}
\usepackage{verbatim}
\usepackage{multirow}
\usepackage{threeparttable}
\usepackage{fixltx2e}
\usepackage[center]{caption} 
\usepackage{lscape}
\usepackage{rotating}
\usepackage{bigdelim}

\begin{document}

\title{Research performance of UNU - A bibliometric analysis of the \\ United Nations University.}
\author{Johannes Stegmann\footnote{Member of the Ernst-Reuter-Gesellschaft der Freunde, F\"orderer und Ehemaligen  der Freien Universit\"at Berlin e.V.}
\footnote{Former (now retired) employee of the Medical Library of the Free University Berlin and the Charit\'e Berlin.}
\footnote{Radebeul, Germany, johannes.stegmann@fu-berlin.de} }
\date{}
\maketitle

\begin{abstract} 
The scientific paper output of the United Nations University (UNU) was bibliometrically analysed. It was 
found that (i) a noticeable continous paper output starts in 1995, (ii) about 65\% of the research papers have been published as international cooperations and 18\% as single-authored papers, (iv) the research papers rank above world average according to Pudovkin-Garfield Percentile Rank Index, and (v) papers' content indicate the wide variety of scientific topics UNU has been and is working on. \\
\textbf{Keywords}: United Nations University, Bibliometry, Pudovkin-Garfield Percentile Rank Index, Content Analysis.
\end{abstract}
\newline 

\section{Introduction} \renewcommand*{\thefootnote}{\fnsymbol{footnote}}
\indent 
\indent The General Assembly of the United Nations formally established the United Nations University - UNU - in 1972 and approved the University Charter and Resolution in December 1973. UNU's academic work was launched in September 1975 
(UNU, 2016a). \\
\indent On UNU's website it is stated: {\em"The mission of the UN University is to contribute, through collaborative research and education, to efforts to resolve the pressing global problems of human survival, development and welfare that are the concern of the United Nations, its Peoples and Member States."} (UNU, 2016b).\\
\indent In addition, article I, paragraph 2 of the University Charter states: {\em "The University shall devote its work to research into the pressing global problems of human survival, development and welfare that are the concern of the United Nations and its agencies, with due attention to the social sciences and the humanities as well as natural sciences, pure and applied."} (UNU, 2016c). \\
\indent In practice, UNU faces three main tasks: (i) to perform interdisciplinary research to cope with imperative global problems; (ii) acting as a "think tank" transferring research results into concrete operation instructions (recommendations) for the United Nations (UN) and its member states; and (iii) to establish and execute postgraduate trainings. All three tasks are, of course, dependent of and intertwined with the necessity of providing and developing appropriate capacities and facilities (UNU, 2016d).\\
\indent It is certainly of interest to analyse UNU's scientific acitivities from a bibliometric point of view, i.e. (i) to investigate UNU's overall productivity and citation performance, (ii) to determine UNU's degree of interdisciplinarity and (international) cooperation, and (iii) to have a look at the thematic subjects UNU is dealing with. \\
\indent The bibliometric investigation presented here uses publication and citation data from the Web of Science database (WoS). It is therefore restricted to papers published in scientific journals indexed in the WoS.

\section{Methods} \renewcommand*{\thefootnote}{\fnsymbol{footnote}}
\subsection{Retrieval, download} \indent
\indent Papers were retrieved and downloaded with all bibliographic data from the Web of Science (WoS) on October 11, 2016, using an appropriate UNU-specific address search profile. \\
\indent For the  analysis of UNU's paper output and its distribution to different document types all retrieved records were included. UNU's' research was analysed from research-relevant papers only, i.e. papers indexed as ARTICLE, PROCEEDINGS PAPER, or REVIEW. 

\subsection{Citation performance} \indent
\indent The citation performance of UNU's papers was measured applying the Percentile Rank Index (PRI) developed by Pudovkin and Garfield (Pudovkin and Garfield, 2009). This version of a percentile rank index was called PG-PRI (Stegmann, 2014). Because no other PRI methods are involved here, "PG-PRI" and "PRI" are used synonymously. \\
\indent For the PG-PRI calculation of an individual paper its citation rank among its "paper peers" must be determined. Therefore, the citation data of all papers of the same document type published in the same journal in the same year are required, i.e. the respected journal-year pairs had to be retrieved and downloaded from the WoS which was done between the 13\textsuperscript{th} and 16\textsuperscript{th} October 2016. PG-PRIs of UNU papers were calculated only for papers of the document type ARTICLE published prior to 2014. \\
\indent The calculation of PG-PRI and determination of the global (expected) average PRI have been described (Pudovkin and Garfield, 2009, Pudovkin et al., 2012, Stegmann, 2014). Briefly, after determination of the position (citation rank) of a paper in the corresponding paper set (the year volume of the respective journal), its PG-PRI value was calculated according to the formula 
\begin{displaymath}
PRI = \frac{N-R+1}{N}*100 
\end{displaymath}
where N is the number of papers in the year set of the journal and R is the citation rank of the paper. R=1 is the top rank (most cited paper) with PRI=100. For determination of the global (expected) average PRI the UNU articles were ordered according to the number of papers published in the corresponding journal-year set. The average PRI was calculated according to the formula
\begin{displaymath}
PRI_{globav} = 50 + \frac{50}{N}.
\end{displaymath}
where N is the number of papers published in the journal-year pair at the median position of the ordered set. \\

\subsection{Research diversity} \indent
\indent To measure the diversity of UNU's research Shannon's entropy formula (also known as {\em Shannon Diversity Index}) (Shannon, 1949, p. 681) was applied to journals, scientific (sub)fields (WoS categories), and keywords of UNU's research papers. The entropy (diversity) {\em H} is calculated according to the formula
\begin{displaymath}
H = -\sum \limits_{i=1}^S p_i\: ln\: p_i
\end{displaymath}
with 
\begin{displaymath}
p_i = \frac{n_i}{N}
\end{displaymath}
where S is the number of species (species richness) (e.g., journals), n\textsubscript{i} the number of individuals in species i (the abundance of species i, i.e. the number of papers in journal i), N the total number of all individuals (papers), and 
p\textsubscript{i} the relative abundance of species (journal) i, calculated as the proportion of individuals (papers) of species (journal) i to the total number of individuals (papers). ln is the natural logarithm. The maximal diversity is equal to the logarithm of S:
\begin{displaymath}
H_{max} = ln\:S
\end{displaymath}

\subsection{Research topics} \indent
\indent To identify the topics of UNU's research the WoS subject categories (record field SC) to which the publishing journals are assigned and the keywords (record field DE: author keywords, record field ID: 
keyword plus\textsuperscript{\textregistered}, assigned during the indexing process) were extracted and analysed. For cluster analysis of keywords the co-word analysis technique described by Callon et al. (1991) was applied. A detailed description of the algorithm can be found in Stegmann and Grohmann (2003). The same thresholds were applied as in Stegmann (2014), i.e. a minimal frequency of 4 ((occurrence of a keyword in at least 4 records), a minimal cosine similarity of 0.2, and a minimal (maximal) cluster size of 3 (10) distinct keywords. To label a cluster (besides its number) the keywords contained in it were ranked according to the product of link strengths (cosine values) and frequency; the keyword with the highest product was selected as cluster name (Stegmann, 2014).

\subsection{Programming, calculations} \indent
\indent Extraction of record field contents, clustering, data analysis, calculations and visualisation were done using homemade programs and scripts for perl (version 5.14.2) and the software package R version 3.2.3 (R Core Team, 2015). All operations were performed on a commercial PC run under Ubuntu version 12.04 LTS.

\section{Results and Discussion} 
\subsection{Output} \indent
\indent Figure 1 shows UN's paper output. A total of 1314 papers were retrieved and downloaded from the WoS. Although UNU's 
academic work began in 1975 (see Introduction), a continous paper output seems to start only after 1995. One reason may be that it certainly takes some time until freshly founded academic institutions are able to do real research apart from administrative and management tasks such as recruitment of staff and purchase of necessary equipment. Another reason might be that initially UNU's researcher perhaps did not use in their paper publications an UNU-specific institutional address with the consequence that those papers cannot be found by an UNU-specific address search in the WoS. On UNU's website publications are listed from 1994 onward ((UNU, 2016e). A quick survey of the years 1994 and 1995 reveals that many of the papers are "research memoranda" and have not been published in regular scientific journal. In the more recent years, however, UNU's paper output is strongly increasing, culminating for now in nearly 200 papers published in 2015 (Figure 1). 

\begin{figure}[H]
\centerline{
\includegraphics[height=8.0cm]{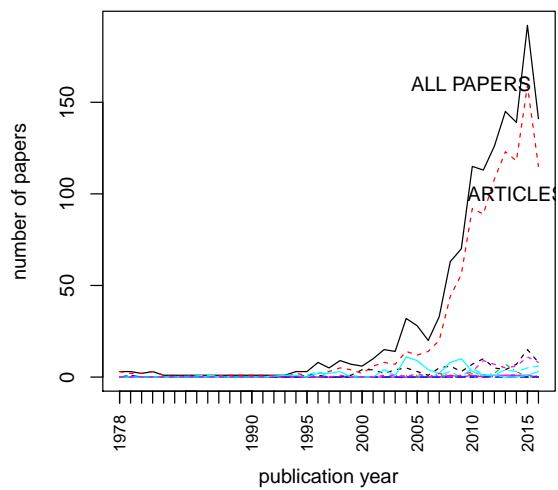}}
\caption{UNU papers published in scientific journals\textsuperscript{*}.}
     *\footnotesize Total no. of papers: 1314
 (77.4\% articles, 7.3\% editorials, 5.19\% proceedings papers, 4.3\% reviews, 2.7\% book reviews, 3.1\% other).
\label{fig:fig01}
\end{figure}

\indent More than three-fourths of UNU's paper output are of document type ARTICLE (Figure 1), indicating a strong research orientation of UNU's institutes.

\subsection{Citation performance (PG-PRI)} \indent
\indent
For the analysis of the international standing of UNU's research articles the percentile rank indexing method of Pudovkin and 
Garfield (2009) was applied (PG-PRI, see Methods). 652 articles published before 2013\footnote{Articles published later were not included because the time span to receive citations would be too short.} were found. They are distributed to 569 journal-year article sets. These sets were retrieved, and the articles of each set were ranked according to the number of cites they had received at the time of retrieval (see Methods). Thus, the citation rank of each of the 652 UNU articles was determined. Figure 2 displays the PG-PRI value of each of the 652 research articles of UNU. The median of UNU's papers as well as the expected global average are also marked. The median of UNU's papers ordered according to the number of papers published in the correponding journal year sets (see Methods) is between 326 and 327, i.e. 326.5. The number of papers in the journal-year sets at position 326 and 327 of the ordered sets are 70 and 71, thus the number of papers at the median position is 70.5. Therefore, the global average PRIglobav = 50.68 (see Methods).

\begin{table}[H]\small
\caption{UNU's research articles prior to 2014: PG-PRI\textsuperscript{*} ranges.}
*\footnotesize PG-PRI: Pudovkin-Garfield Percentile Rank Index.
%
%
\centering
\begin{tabular}{lcc}
\noalign{\smallskip}
\hline\noalign{\smallskip}
PRI range &  No. of papers &  \% of total (652 papers)    \\
\noalign{\smallskip} 
\noalign{\smallskip}\hline\noalign{\smallskip}
PRI = 100 & 17 & 2.7  \\
\noalign{\smallskip} 
PRI $\ge$ 99 & 49 & 7.5 \\
\noalign{\smallskip} 
PRI $\ge$ 90 & 102 & 15.6  \\
\noalign{\smallskip} 
PRI $\ge$ 75 & 206 & 31.6   \\
\noalign{\smallskip} 
PRI $\ge$ 50.7 & 363 & 55.7   \\
\noalign{\smallskip} 
\noalign{\smallskip}\hline
\end{tabular}
\end{table} 
\begin{figure}[H]
  \centerline{
  \includegraphics[height=8.0cm]{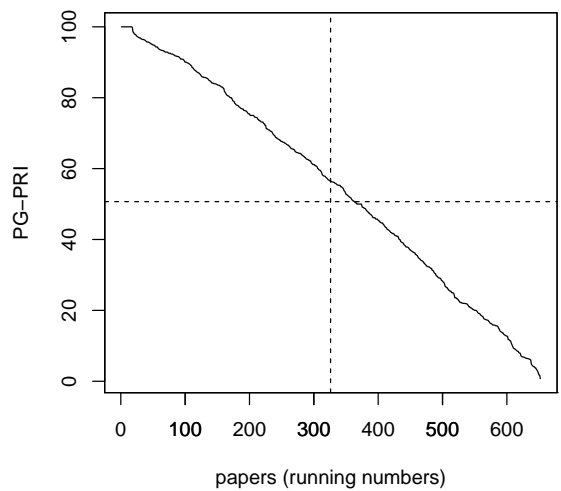}}
    \caption{Percentile Rank Indexes of UNU's research articles published prior to 2014\textsuperscript{*}.}
   *\footnotesize Total no. of papers: 652.\newline
     PG-PRI: Pudovkin-Garfield Percentile Rank Index.\newline
     Vertical dashed line: median of papers.\newline
     Horizontal dashed line: expected global mean PRI.
   \label{fig:fig02}
\end{figure}

\indent 55.7\% of UNU's papers have an PRI value higher than the global average (Figure 2, Table 1). The average PRI of UNU's articles is 55.1 ((sum of PRI values divided by the number of papers), likewise considerably higher than the global mean. Table 1 lists some PRI ranges. Nearly 56\% of the papers have PRI values above the global mean. \\
\indent From the data in Figure 2 and Table 1 it is concluded that UNU's research perform well above the average of comparable world research.

\subsection{Research output} \indent
\indent In this section, UNU's research output is analysed, i.e. all 1141 UNU papers which were indexed as ARTICLE, PROCEEDINGS PAPER, or REVIEW and could be retrieved and downloaded from the WoS on October 11, 2016 (see Methods).

\subsubsection{Authorship and cooperation} \indent
\indent Figure 3 displays different authorship and cooperation types. The majority of UNU's research papers is multi-authored and published as national (17\%) or international cooperations (65\%). About 18\% of the papers have only one author. This value seems to be rather high because one should expect all UNU research papers to be multi-authored, cooperation being a dedicated goal of UNU's activities (see Introduction). To give an example, only 4.1\% of the research papers of he University Centre in Svalbard are single-authored (Stegmann, 2014). A closer look at UNU's single-authored papers, however, did not reveal any peculiarities, such as predominance of specific topics or publication years (data not shown). 

\begin{figure}[H]
\centerline{
\includegraphics[height=8.0cm]{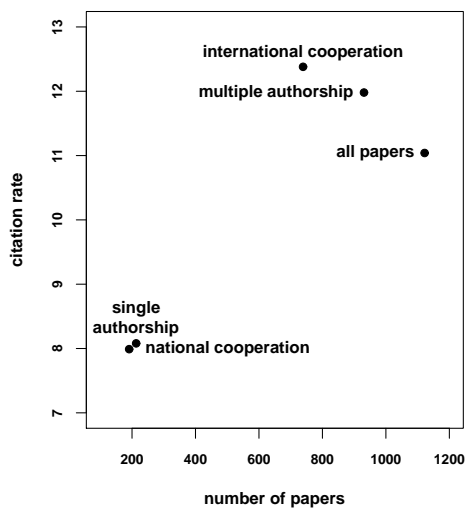}}
\caption{UNU research papers: different types of authorship and cooperation\textsuperscript{*}.}
  *\footnotesize All (distinct) papers: 1141 (100\%). \newline
   Single authorship: 18.7\%.\newline
   Multiple authorship: 81.6\%.\newline
   National (intra- and extramural) cooperation: 16.8\%.\newline
   International cooperation: 64.8\%.
\label{fig:fig03}
\end{figure}

\newpage

\indent For completeness, Figure 3 displays also the citation rates of the different paper sets although recent papers certainly are not (yet) or very little cited. But it seems to be clear that papers published as international cooperation (i.e. published by authors from different countries) have in general higher citation rates, a well known phenomenon in bibliometrics (Narin et al., 1991, Katz and Hicks, 1997, Glaenzel, 2001, Stegmann and Grohmann, 2005, 2006).

\begin{table}[H]\small
\caption{UNU cities: research productivity\textsuperscript{*}.}
*\footnotesize Minimum number of papers: 1\% of output.
%
%
\centering
\begin{tabular}{lcc}
\noalign{\smallskip}
\hline\noalign{\smallskip}
City &  No. of papers &  \% of total  \\
     &                &   (1141 distinct papers) \\ \noalign{\smallskip} 
\noalign{\smallskip}\hline\noalign{\smallskip}
BONN & 136 & 11.9  \\
\noalign{\smallskip}
BRUGGE & 36 & 3.2 \\
\noalign{\smallskip} 
ENSCHEDE & 12 & 1.1   \\
\noalign{\smallskip} 
HAMILTON & 71 & 6.2   \\
\noalign{\smallskip} 
HELSINKI & 156 & 13.7   \\
\noalign{\smallskip} 
KUALA LUMPUR & 65 & 5.7   \\
\noalign{\smallskip} 
MAASTRICHT & 251 & 22.0   \\
\noalign{\smallskip} 
MACAO & 42 & 3.7  \\
\noalign{\smallskip} 
REYKJAVIK & 16 & 1.4   \\
\noalign{\smallskip} 
TOKYO & 207 & 18.1   \\
\noalign{\smallskip} 
YOKOHAMA & 99 & 8.7   \\
\noalign{\smallskip} 
\noalign{\smallskip}\hline
\end{tabular}
\end{table} 

\subsubsection{UNU cities} \indent
\indent UNU's research is performed within the framework of several institutes and programmes located around the world. The current state of UNU's activities is published on UNU's website (UNU, 2016f). UNU seems to be in permanent developmental
and (re-)structuring processes, including establishment of new, re-start or re-orientation and even close-down of existing institutes and programmes (UNU, 2016g). Thus, a bibliometric analysis may only partially reflect the present situation of UNU's research infrastructure. Initially, it was planned to relate the papers to UNU's labeled research activities, like "UNU-WIDER" or "UNU-MERIT" (see UNU, 2016f). However, numerous papers carry as UNU-specific address strings like "UNITED NATIONS UNIV", followed by a city name but not a definite institute or programm designation. Thus, Table 2 relates UNU's research papers to "UNU cities", i.e. cities which harbor UNU institutes and/or programmes. A threshold of one percent of the research papers was applied. Eleven cities were found. Cities not included but being mentioned on UNU's website (UNU, 2016f) may have become an UNU location only recently. Vice versa, cities mentioned in Table 2 but not appearing on UNU's website (UNU, 2016f) harbor either cooperating (with UNU) institutions or hosted moved or closed-down UNU institutions. In fact, UNU's website obviously reflect only the present state of UNU activities. Former activities can be found in UNU's Annual Reports (UNU, 2016h).

\subsubsection{Research diversity} \indent
\indent UNU's research seems to be fairly diversified. The 1141 research papers have been published in 527 journals in 88 WoS categories (scientific subfields). The average is about 2 papers per journal or 13 papers per subfield. The total number of keywords is 5232, i.e. on the average, each  paper was classified by 4-5 keywords. Table 3 shows the diversity indexes calculated for journals, categories, and keywords applying Shannon's diversity formula (see Methods). The diversity indexes of journals and keywords nearly reach the maximum. The reduction of the number of keywords by the clustering process does not significantly change their diversity; it is even slightly higher (see Table 3). \\

\indent Table 4 lists the 18 journals with 10 or more UNU papers. To show the international standing of the journals their actual (2015) impact factor is mentioned. From Table 4 it is clear that none of the journals is predominant, hence the large diversity (see Table 3). \\
\indent Table 5 list the 10 WoS categories with at least 50 UNU papers. Here the two categories {\em ENVIRONMENTAL SCIENCES ECOLOGY} and {\em BUSINESS ECONOMICS} predominate with (both together) more than 50\% of the UNU's research papers, hence the subfield diversity is not so pronounced (see\hspace{0.1cm} Table 3). \\

\begin{table}\small
\caption{UNU's research papers: diversity\textsuperscript{*}.}
*\footnotesize {\em H}: diversity index, {\em H\textsubscript{max}}: maximal diversity (see Methods)
%
%
\centering
\begin{tabular}{lrccc}
\noalign{\smallskip}
\hline\noalign{\smallskip}
type &  no. &  {\em H} &  \% of {\em H\textsubscript{max}} & {\em H\textsubscript{max}}  \\
\noalign{\smallskip}\hline\noalign{\smallskip}
journals & 527 & 5.83 & 93 & 6.27  \\
\noalign{\smallskip}
categories & 88 & 3.15 & 70 & 4.48 \\
\noalign{\smallskip} 
keywords  & 5232 & 7.98 & 93 & 8.56   \\
 (total)  &      &      &    &   \\
\noalign{\smallskip} 
keywords  & 341 & 5.53 & 95 & 5.83   \\
(cluster terms) &  &    &    &  \\
\noalign{\smallskip} 
\noalign{\smallskip}\hline
\end{tabular}
\end{table} 

\begin{table*}\small
\caption{UNU's research papers: publishing journals (top 18)\textsuperscript{*}.}
*\footnotesize Journals with at least 10 UNU papers. \newline
  IF 2015: Impact factor, derived from the {\em Journal Citation Reports\textsuperscript{\textregistered}}.
%
%
\centering
\begin{tabular}{lccc}
\noalign{\smallskip}
\hline\noalign{\smallskip}
Journal  & IF 2014 &  No. of papers &  \% of total  \\
         &         &                &   (1141 distinct papers) \\ \noalign{\smallskip} 
\noalign{\smallskip}\hline\noalign{\smallskip}
SUSTAINABILITY SCIENCE & 2.49 & 26 & 2.3 \\
\noalign{\smallskip} 
RESEARCH POLICY  & 3.47  & 22  & 1.9  \\
\noalign{\smallskip} 
JOURNAL OF CLEANER PRODUCTION & 4.96  & 20  & 1.8  \\
\noalign{\smallskip} 
TECHNOLOGICAL FORECASTING AND SOCIAL CHANGE & 2.68  & 17  & 1.5  \\
\noalign{\smallskip} 
PLOS ONE & 3.06  & 17  & 1.5 \\
\noalign{\smallskip} 
WORLD DEVELOPMENT & 2.44  & 17  & 1.5  \\
\noalign{\smallskip} 
ENERGY POLICY & 3.05  & 13  & 1.1  \\
\noalign{\smallskip} 
ENVIRONMENTAL SCIENCE \& POLICY & 2.97  & 12  & 1.1  \\
\noalign{\smallskip} 
AFRICAN DEVELOPMENT REVIEW & 0.72  & 12  & 1.1 \\
\noalign{\smallskip} 
SUSTAINABILITY & 1.34  & 11  & 0.9  \\
\noalign{\smallskip} 
NATURAL HAZARDS & 2.05  & 11 & 0.9  \\
\noalign{\smallskip} 
SCIENCE OF THE TOTAL ENVIRONMENT & 3.98 & 11  & 0.9   \\
\noalign{\smallskip} 
LAND USE POLICY & 2.77  & 11  &0.9  \\
\noalign{\smallskip} 
SMALL BUSINESS ECONOMICS & 1.80  & 10  & 0.9  \\
\noalign{\smallskip} 
JOURNAL OF AFRICAN ECONOMIES & 0.62  & 10  & 0.9   \\
\noalign{\smallskip} 
CURRENT OPINION IN ENVIRONMENTAL SUSTAINABILITY & 4.66  & 10 & 0.9  \\
\noalign{\smallskip} 
CLIMATIC CHANGE & 3.34  & 10  & 0.9  \\
\noalign{\smallskip} 
JOURNAL OF DEVELOPMENT STUDIES  & 0.90  & 10  & 0.9 \\
\noalign{\smallskip} 
\noalign{\smallskip}\hline
\end{tabular}
\end{table*} 

\begin{table*}\small
\caption{UNU's research papers: WoS categories (top 10)\textsuperscript{*}.}
*{\footnotesize Subfields with at least 50 UNU papers.} \newline
%
%
\centering
\begin{tabular}{lcc}
\noalign{\smallskip}
\hline\noalign{\smallskip}
WoS category  &  No. of papers &  \% of total  \\
              &                &   (1141 distinct papers) \\ \noalign{\smallskip} 
\noalign{\smallskip}\hline\noalign{\smallskip}
ENVIRONMENTAL SCIENCES ECOLOGY & 327 & 28.7 \\
\noalign{\smallskip} 
BUSINESS ECONOMICS  & 322  & 28.2  \\
\noalign{\smallskip} 
PUBLIC ADMINISTRATION  & 145  & 12.7  \\
\noalign{\smallskip} 
SCIENCE TECHNOLOGY OTHER TOPICS & 121  & 10.6  \\
\noalign{\smallskip} 
ENGINEERING & 92  & 8.1 \\
\noalign{\smallskip} 
WATER RESOURCES & 80  & 7.0  \\
\noalign{\smallskip} 
GEOLOGY & 54  & 4.7  \\
\noalign{\smallskip} 
AGRICULTURE  & 51  & 4.5  \\
\noalign{\smallskip} 
INTERNATIONAL RELATIONS & 51  & 4.5 \\
\noalign{\smallskip} 
METEOROLOGY ATMOSPHERIC SCIENCES & 50  & 4.4  \\
\noalign{\smallskip} 
\noalign{\smallskip}\hline
\end{tabular}
\end{table*} 
%
%

\begin{figure*}[hbtp]\small
  \centerline{
 \includegraphics[width=\textwidth]{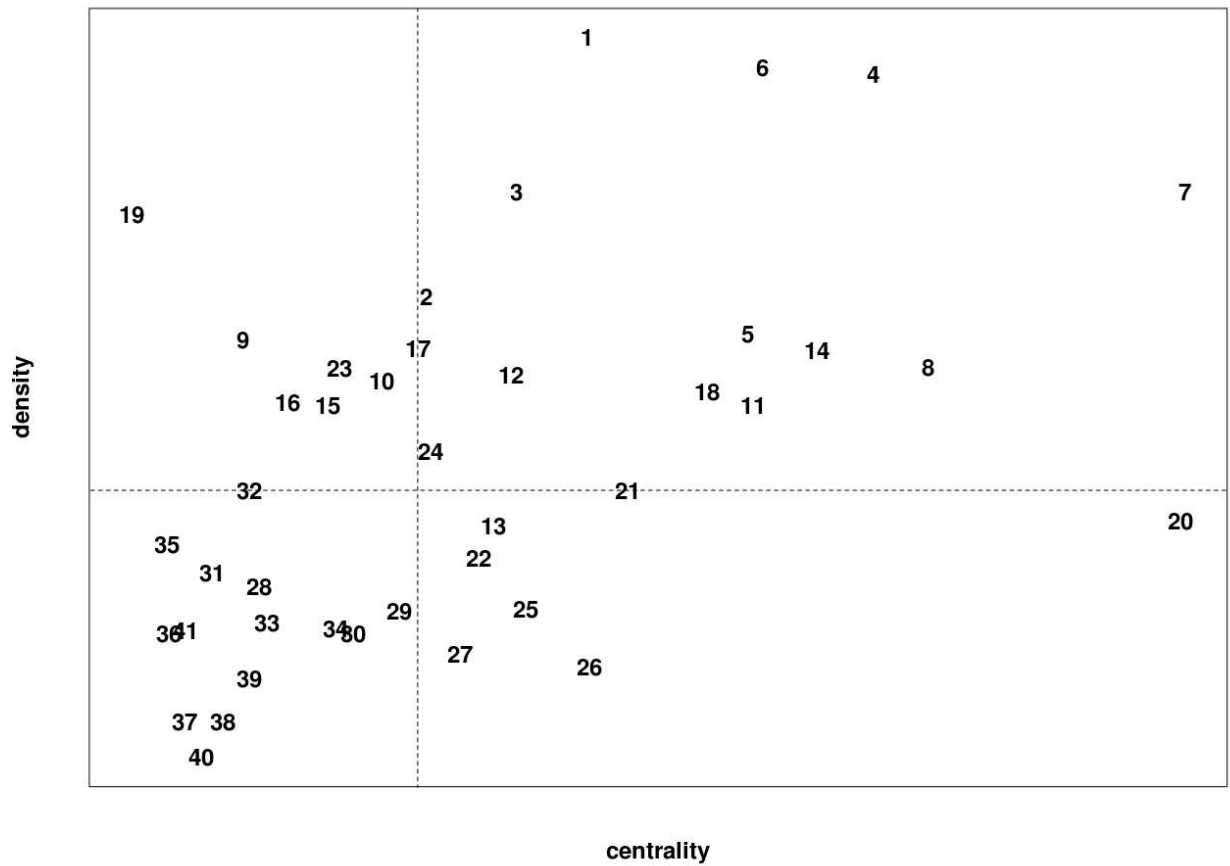}}
    \caption{UNU's research papers: centrality-density diagram of keywords clusters\textsuperscript{*}.}
     *\footnotesize Cluster numbers are centered at cluster positions.\newline
       Centrality increases from left to right, Density increases bottom-up.\newline
       Vertical dashed line: median centrality.\newline
       Horizontal dashed line: median density.
   \label{fig:fig04}
\end{figure*}

\begin{table*}
\caption{UNU's research: clusters and terms\textsuperscript{*} positioned in the upper left quadrant\textsuperscript{**} of Figure 4.} 
*\footnotesize{\em Italic terms}: cluster label terms (see Methods). \newline
**\footnotesize above median density and below median centrality. \newline
The terms of the clusters in this quadrant occur in 290 distinct papers.

%
\begin{tabular}{ccl}
\noalign{\smallskip}
\hline\noalign{\smallskip} 
cluster  & no. of &  \hspace{5cm} keywords  \\
   no.   & papers &           \\
\noalign{\smallskip} 
\noalign{\smallskip}\hline\noalign{\smallskip}
9 & 37 & \footnotesize{\em REFINEMENT}; VERIFICATION; TANZANIA; GREENHOUSE GAS EMISSIONS; SEMANTICS; \\
  &    &  \footnotesize UML; BIOFUELS; FORMAL METHODS; LIFE CYCLE ASSESSMENT; MITIGATION \\ 
\noalign{\smallskip}
10 & 34 & \footnotesize{\em FOREIGN DIRECT INVESTMENT}; DOMESTIC FIRMS; KNOWLEDGE SPILLOVERS; \\
  &    &  \footnotesize PATENT CITATIONS; FDI; DIRECT INVESTMENT; INNOVATION SYSTEMS; OFFSHORING; \\
  &    &  \footnotesize PROTECTION; AGGLOMERATION \\ 
\noalign{\smallskip}
15 & 50 & \footnotesize{\em ENTREPRENEURSHIP}; ECONOMIC DEVELOPMENT; BARRIERS; START UPS; RETURNS; \\
  &    & \footnotesize DEVELOPMENT; SECTOR; CHOICE; UGANDA; INCOME \\ 
\noalign{\smallskip}
16 & 81 & \footnotesize{\em INDIA}; CITY; DELHI; DECOMPOSITION; INDICATORS; INDEX; GHG EMISSIONS; \\
  &    & \footnotesize CO BENEFITS; INCOME INEQUALITY; PATTERNS  \\ 
\noalign{\smallskip}
17 & 48 & \footnotesize{\em LAND DEGRADATION}; DESERTIFICATION; DRYLANDS; SUSTAINABLE LAND MANAGEMENT; \\
  &    & \footnotesize EXPERIENCES; SCIENCE POLICY INTERFACE; ECONOMIC VALUATION; SYSTEM; \\
  &    & \footnotesize NATURAL RESOURCE MANAGEMENT; DEGRADATION \\ 
\noalign{\smallskip}
19 & 29 & \footnotesize{\em CONFLICT}; CIVIL WAR; DETERMINANTS \\ 
\noalign{\smallskip}
23 & 50 & \footnotesize{\em EMPIRICAL ANALYSIS}; LATIN AMERICA; SIMULATION; GOVERNMENT; CERTIFICATION; \\
  &    & \footnotesize ETHIOPIA; VARIABILITY; CORRUPTION; INSTITUTIONS; INSURANCE \\ 
\noalign{\smallskip}
\noalign{\smallskip}\hline
\noalign{\smallskip}
\end{tabular} 
\end{table*}

\begin{table*}
\caption{UNU's research: clusters and terms\textsuperscript{*} positioned in the upper right quadrant\textsuperscript{**} of Figure 4.} 
*\footnotesize{\em Italic terms}: cluster label terms (see Methods). \newline 
**\footnotesize above median density and above median centrality. \newline
The terms of the clusters in this quadrant occur in 555 distinct papers.

\begin{tabular}{ccl}
\noalign{\smallskip}
\hline\noalign{\smallskip}
cluster  & no. of &  \hspace{5cm} keywords  \\
   no.   & papers &           \\
\noalign{\smallskip} 
\noalign{\smallskip}\hline\noalign{\smallskip}
1 & 37 & \footnotesize{\em EXPOSURE}; PREVALENCE; INDOOR ENVIRONMENTS; QUANTITATIVE PCR; MORTALITY; \\
  &    & \footnotesize ASSOCIATION; CANCER; CARE; MASS SPECTROMETRY; RISK FACTORS \\ 
\noalign{\smallskip}  
2 & 61 & \footnotesize{\em ECONOMIC GROWTH}; FIRM GROWTH; EDUCATION; COTE DIVOIRE; QUANTILE REGRESSION; \\
  &    & \footnotesize VOLATILITY; ENDOGENOUS GROWTH; SIZE; BASIC NEEDS; TIME SERIES \\ 
\noalign{\smallskip}  
3 & 137 & \footnotesize{\em CLIMATE CHANGE}; VULNERABILITY; ADAPTATION; TSUNAMI; INDONESIA; EARTHQUAKE; \\
  &    & \footnotesize DISASTER RISK REDUCTION; NATURAL HAZARDS; SUMATRA; SRI LANKA  \\ 
\noalign{\smallskip}
4 & 32 & \footnotesize{\em REMITTANCES}; ACCUMULATION; LABOR MIGRATION; AID; CORAL REEFS; RECRUITMENT; \\
  &    & \footnotesize COASTAL DEFENSE STRUCTURES; REFUGEES; ENVIRONMENTAL DEGRADATION;   \\ 
  &    & \footnotesize CONSEQUENCES \\
\noalign{\smallskip}
5 & 51 & \footnotesize{\em CONSUMPTION}; PERSIAN GULF; EMISSIONS; DEFORESTATION; ARABIAN GULF; \\
  &    & \footnotesize ENERGY EFFICIENCY; DESIGN; ASSEMBLAGES; FISH; COMMUNITY STRUCTURE \\ 
\noalign{\smallskip}
6 & 45 & \footnotesize{{\em CAMBODIA}; CONTAMINATION; RISK ASSESSMENT; WEST BENGAL; BANGLADESH;} \\
  &    & \footnotesize{GROUNDWATER; ORGANOCHLORINE PESTICIDES; ARSENIC; ECOLOGICAL RISK ASSESSMENT;} \\
  &    & \footnotesize{FOOD CONSUMPTION} \\ 
\noalign{\smallskip}
7 & 106 & \footnotesize{\em INNOVATION}; KNOWLEDGE; TECHNOLOGY; STRATEGIC ALLIANCES; SOCIAL STRUCTURE; \\
  &    & \footnotesize CAPABILITIES; PHARMACEUTICAL INDUSTRY; EMBEDDEDNESS; INTERFIRM NETWORKS;  \\ 
  &    & \footnotesize EVOLUTION \\
\noalign{\smallskip}
8 & 49 & \footnotesize{\em DRINKING WATER}; POLLUTION; TECHNOLOGICAL INNOVATION; RICE; FARMERS; \\
  &    & \footnotesize MEKONG DELTA; RAINWATER; BENEFITS; REDD; ADOPTION \\ 
\noalign{\smallskip}
11 & 55 & \footnotesize{\em LAND USE}; WATER QUALITY; RIVER; PERSISTENT ORGANIC POLLUTANTS; SOIL; \\
  &    & \footnotesize SEDIMENTS; RUNOFF; FATE; CATCHMENT; EROSION \\ 
\noalign{\smallskip}
12 & 38 & \footnotesize{\em MARINE PROTECTED AREAS}; ECOSYSTEMS; RECOVERY; ECONOMY; SOCIETY; \\
  &    & \footnotesize CORAL REEF; FIELD; TRANSFORMATION; GREAT BARRIER REEF; CARBON \\ 
\noalign{\smallskip}
14 & 45 & \footnotesize{\em RESEARCH AND DEVELOPMENT}; ABSORPTIVE CAPACITY; COMPETITIVE ADVANTAGE; \\
  &    & \footnotesize R\&D COLLABORATION; OPEN INNOVATION; DYNAMIC CAPABILITIES; COMPLEMENTARITY; \\
  &    & \footnotesize INTERNATIONAL JOINT VENTURES; PRODUCT INNOVATION; TECHNOLOGY TRANSFER \\ 
\noalign{\smallskip}
18 & 61 & \footnotesize{\em HEALTH}; ENERGY; DECISION MAKING; KENYA; TECHNOLOGIES; \\
  &    & \footnotesize INDUSTRIAL ECOLOGY; GEOGRAPHY; COLLECTIVE ACTION; ELECTRICITY; COSTS \\ 
\noalign{\smallskip}
24 & 71 & \footnotesize{\em BIODIVERSITY}; CONSERVATION; ECOSYSTEM SERVICES; IMPLEMENTATION; AREAS; \\
  &     & \footnotesize URBAN AREAS; FISHERIES MANAGEMENT; COMANAGEMENT; POLICIES;  \\ 
  &     & \footnotesize TRADITIONAL KNOWLEDGE \\
\noalign{\smallskip}
\noalign{\smallskip}\hline
\end{tabular} 
\end{table*}

\begin{table*}
\caption{UNU's research: clusters and terms\textsuperscript{*} positioned in the lower left quadrant\textsuperscript{**} of Figure 4.}
*\footnotesize{\em Italic terms}: cluster label terms (see Methods). \newline
**\footnotesize below median density and below median centrality. \newline
The terms of the clusters in this quadrant occur in 456 distinct papers.
%
\begin{tabular}{ccl}
\noalign{\smallskip}
\hline\noalign{\smallskip}
cluster  & no. of &  \hspace{5cm} keywords  \\
   no.   & papers &           \\
\noalign{\smallskip} 
\noalign{\smallskip}\hline\noalign{\smallskip}
28 & 47 & \footnotesize{\em UNEMPLOYMENT}; INEQUALITY; WAGES; EMPLOYMENT; EXPERIENCE; IMMIGRATION; THAILAND \\ 
\noalign{\smallskip}
29 & 60 & \footnotesize{\em MIGRATION}; FOOD SECURITY; FOOD; DISPLACEMENT; RAINFALL VARIABILITY; DISASTER; \\
  &    & \footnotesize COOPERATION; NATURAL DISASTERS; DIFFUSION; CLIMATE CHANGE ADAPTATION  \\ 
\noalign{\smallskip}
30 & 30 & \footnotesize{\em UK}; OUTCOMES; INTELLECTUAL PROPERTY RIGHTS; QUALITY OF LIFE; \\
  &    & \footnotesize COST EFFECTIVENESS; COST; TRANSPORT \\ 
\noalign{\smallskip}
31 & 23 & \footnotesize{\em ENVIRONMENTAL CHANGE}; WORLD; GLOBALIZATION \\ 
\noalign{\smallskip}
32 & 30 & \footnotesize{\em PUBLICATIONS}; NANOTECHNOLOGY; PATENTS; BIBLIOMETRICS; ASIA \\ 
\noalign{\smallskip}
33 & 61 & \footnotesize{\em FRAMEWORK}; CHALLENGES; LESSONS; RESPONSES; FOREIGN AID; DISASTERS; \\
  &    & \footnotesize FRAGILE STATES; SUSTAINABILITY SCIENCE; HAZARDS; SOUTHERN AFRICA \\ 
\noalign{\smallskip}
34 & 150 & \footnotesize{\em POVERTY}; GROWTH; SOUTH AFRICA; TRADE; CGE MODEL; MARKET; POOR; \\
   &    & \footnotesize EXPORTS; AGRICULTURE; MOZAMBIQUE \\ 
\noalign{\smallskip}
35 & 52 & \footnotesize{{\em PRODUCTIVITY}; R\&D; MANUFACTURING FIRMS; TECHNOLOGICAL CHANGE; SMES} \\ 
\noalign{\smallskip}
36 & 43 & \footnotesize{{\em URBAN}; POPULATION; EPIDEMIOLOGY; SCALE; REMOTE SENSING; ECOLOGY} \\ 
\noalign{\smallskip}
37 & 27 & \footnotesize{\em EFFICIENCY}; HOUSEHOLDS; ALLOCATION \\ 
\noalign{\smallskip}
38 & 13 & \footnotesize{\em WELFARE}; NIGERIA; EUROPEAN UNION \\ 
\noalign{\smallskip}
39 & 16 & \footnotesize{\em PARTICIPATION}; TOXICITY; AGGREGATION; STRESS \\ 
\noalign{\smallskip}
40 & 14 & \footnotesize{\em LAND}; WAR; LAND ADMINISTRATION \\ 
\noalign{\smallskip}
41 & 29 & \footnotesize{\em TRANSITIONS}; PERSPECTIVE; TRANSITION \\ 
\noalign{\smallskip}
\noalign{\smallskip}\hline
\noalign{\smallskip}
\end{tabular} 
\end{table*}

\begin{table*}
\caption{UNU's research: clusters and terms\textsuperscript{*} positioned in the lower right quadrant\textsuperscript{**} of Figure 4.}
*\footnotesize{\em Italic terms}: cluster label terms (see Methods). \newline
**\footnotesize below median density and above median centrality. \newline
The terms of the clusters in this quadrant occur in 248 distinct papers.

\begin{tabular}{ccl}
\noalign{\smallskip}
\hline\noalign{\smallskip}
cluster  & no. of &  \hspace{5cm} keywords  \\
   no.   & papers &           \\
\noalign{\smallskip} 
\noalign{\smallskip}\hline\noalign{\smallskip}
13 & 36 & \footnotesize{\em DISEASE}; SANITATION; CARBON DIOXIDE EMISSIONS; EU; USA; ASIA PACIFIC; \\
  &    & \footnotesize REGIONAL INTEGRATION; DIARRHEA; INFECTIOUS DISEASES; PREVENTION \\ 
\noalign{\smallskip}
20 & 45 & \footnotesize{\em NETWORKS}; BIOTECHNOLOGY; INTERORGANIZATIONAL COLLABORATION; COLLABORATION; \\
  &    & \footnotesize FIRM PERFORMANCE; RESOURCE BASED VIEW; ALLIANCES; KNOWLEDGE TRANSFER; \\
  &    & \footnotesize STRATEGY; SMALL WORLD \\ 
\noalign{\smallskip}
21 & 56 & \footnotesize{\em RESILIENCE}; SOCIAL ECOLOGICAL SYSTEMS; ADAPTIVE CAPACITY; SEDIMENT; AMAZON; \\
  &    & \footnotesize WEST AFRICA; RISK MANAGEMENT; ADAPTIVE GOVERNANCE; CIRCULATION; CLIMATE VARIABILITY \\ 
\noalign{\smallskip}
22 & 35 & \footnotesize{\em PANEL DATA MODELS}; FIRM; SEARCH; ENVIRONMENTAL MANAGEMENT; ENTERPRISES; CLUSTER; \\
  &    & \footnotesize CLEAN DEVELOPMENT MECHANISM; DEVELOPING COUNTRY; GEOTHERMAL ENERGY; DROUGHT \\ 
\noalign{\smallskip}
25 & 79 & \footnotesize{\em PERFORMANCE}; PANEL DATA; MODELS; SERVICES; SPILLOVERS; UNIVERSITIES; \\
  &    & \footnotesize INFORMATION TECHNOLOGY; PRODUCTIVITY GROWTH; SELECTION; HETEROGENEITY \\ 
\noalign{\smallskip}
26 & 36 & \footnotesize{\em TEMPERATURE}; CALIBRATION; FOREST; RIVER BASIN; ORGANIZATIONS; \\
  &    & \footnotesize AIR POLLUTION; EVALUATION; LEARNING; YIELD; FLOODS \\ 
\noalign{\smallskip}
27 & 50 & \footnotesize{\em DYNAMICS}; AREA; VEGETATION; SOIL EROSION; HISTORY; WATER RESOURCES; \\
  &    & \footnotesize IRRIGATION; INCENTIVES; INTENSITY \\ 
\noalign{\smallskip}
\noalign{\smallskip}\hline
\end{tabular} 
\end{table*}
%
%

\subsubsection{Centrality-density diagram} \indent
\indent  Whereas the keyword diversity seems to be fairly high, one must consider that WoS' keyword system is not a controlled vocabulary like MESH\footnote{MESH = Medical Subject Headings is MEDLINE's hierarchichal thesaurus.} So, there may be different terms representing the same concept(s)\footnote{see also some cluster terms in Tables 6-9}. Therefore, a cluster analysis of the keyword set was performed. In principle, a cluster analysis should assemble terms with mutual tight connections into the same clusters which then represent concise concepts. Applying the thresholds mentioned in the {\em Methods} section during the clustering process reduced the number of keywords to 341 which were collated to 41 clusters, each containing 3 to 10 terms (see Methods), representing 819 distinct papers. The strong reduction of the number of keywords does not affect their diversity (see above and Table 3) which also indicates a diversified research landscape of UNU. \\
\indent Figure 4 displays the clusters in the form of a centrality-density diagram (Callon et al., 1991). Centrality means the strength of links between clusters, density means the strength of the links between cluster members. The clusters are distributed to four quadrants by the medians of density and centrality. The clusters below median density have less and/or weaker internal links, the clusters to the left of median centrality have less and/or weaker external links. \\
\indent Theoretically, the diagram displays in its right half "central" topics. i.e. subjects with strong or many links to other topics (clusters). In the upper half the topics are "dense", i.e. the keywords constituting a topic are mutually strongly linked but do not have necessarily many or strong links outside the cluster. In the left and lower parts of the diagram subjects developing to more centrality and/or density are displayed. \\
\indent This clustering technique was shaped in order to visualise the different developmental stages of well defined research topics (Callon et al., 1991). But although we here deal with UNU's rather diverse research areas (see above and UNU, 2016b), the centrality-density clustering method scatters UNU's research topics over the whole diagram. The clusters of Figure 4 represent different research topics, not only subtopics of a main research theme. 

\subsubsection{Research topics} \indent
\indent Tables 6, 7, 8 and 9 list all clusters with their keywords. Table 6 and Table 7 show the clusters of the upper half (containing the two quadrants above median density) of Figure 4, Table 8 and Table 9 of the lower half (the two quadrants below or equal to median density). Table 6 and Table 8 display the clusters of the left half (the two quadrants below or equal to median centrality) of Figure 4, Table 7 and Table 9 of the right half (the two quadrants above median centrality). \\
\indent Surveying the clusters quickly reveals that in most clusters terms (keywords) are assembled which are judged intuitively as belonging together and constituing a research topic. Tables 6-9 also indicate the number of distinct papers in which the terms of a cluster occur. Likewise, the number of papers containing the terms of all clusters of each of the four quadrants formed by the median lines of density and centrality (Figure 4) are given (see Tables 6, 7, 8, 9). \\
\indent The clusters listed in Table 6 deal with gas emissions, economic subjects, land (mis)management and politics. Table 7 lists a bundle of different topics, ranging from cancer to ecology, biodiversity, water quality and climate change. The clusters shown in Table 8 deal with (un)employment,poverty, migration, sustainability, and globalisation. Finally, topics displayed in Table 9 deal with social ecology, models, disease issues, air pollution, soil erosion, water resources. \\
\indent All of the topics listed in Tables 6-9 obviously mark and identify (global) problem areas which are not only worthwhile to be worked on but also demand, on one hand, immediate (political) actions in order to prevent irreversible damages of existing social and ecologic systems, and on the other hand, require amply research efforts for sustainable problem solutions. \\
\indent From the research diversity results (see above) and the journal, subfield and keyword data shown in Table 3-9 and Figure 4 one certainly may draw the conclusion that UNU is engaged in research on sustainability, environment, economics, public administration, and many other topics. The question arises, however, whether UNU has a sustained basis for adequate research in so many fields, a problem addressed in UNU's strategic plan 2015-2019 (UNU, 2016g).

\section{Conclusion} \indent
\indent The bibliometric analysis presented in this paper has asserted that the United Nations University (UNU) performs high-level research well above world average in a wide variety of research subjects dealing with global problem areas challenging for immediate and sustainable solutions. It has also be found that the degree of UNU's (international) research cooperations - mirrored in the proportion of single- and multi-authored research papers - is certainly developable.

\newpage

\end{document}